 \documentclass[aps,pra, onecolumn,12 pts,a4paper,showpacs,nofootinbib]{revtex4}
\usepackage{amssymb,amsmath,amsfonts,graphicx,color}
\usepackage{bm}

\begin{document}

\title{Dynamics of self-gravitating systems : Variations on a theme by Michel H\'enon}

\author{Yves Pomeau$^1$ and Martine Le Berre$^2$}

\affiliation{$^1$Department of Mathematics, University of Arizona, Tucson, USA}
\affiliation{$^2$Institut des Sciences Mol\'eculaires d'Orsay ISMO
- CNRS, Universit\'e Paris-Sud, Bat. 210, 91405 Orsay Cedex, France.}

\date{\today }
\begin{abstract}
\textbf{Abstract}
In this contribution to the volume in memoriam of Michel H\'enon, we thought appropriate to look at his early scientific work devoted to the dynamics of large assemblies of interacting masses. He predicted in his PhD thesis that, in such a system, first a collapse of mass occurs at the center and that later binaries stars are formed there. Henceforth, the negative energy of binding of pairs becomes a source of positive energy for the rest of the cluster which evaporate because of that. We examine under what conditions such a singularity can occur, and what could happen afterwards. We hope to show that this fascinating problem of evolution of self-gravitating clusters keeps its interest after the many years passed since H\'enon thesis, and is still worth discussing now.

\end{abstract}

\maketitle
\section{Introduction}

%On a less urgent matter, the proceeding of the conference will be published
%by "edition Hermann"  (http://www.editions-hermann.fr/)
%in a volume dedicated entirely to Michel Henon. You are invited to make
%contributions, which are not required to be original research material.
%You are invited to give your contributions by the end of February 2014.
%French speakers can decide the language of their contribution. Many of you
%have already accepted; would those who have not done so yet, please let us know ?

This contribution to the volume dedicated to Michel H\'enon is a comment and amplification of a theme he introduced in his PhD thesis, defended in 1961 at the University of Paris  \cite{theseMH}. Evry Schatzman was his advisor, and Andr\'e Danjon (by this time, an old man, born in 1880) chaired the committee.
Shortly afterwards, Michel H\'enon went to Princeton University, being invited by Lyman Spitzer. There he invented the famous "H\'enon-Heiles" model, introduced as a student 3-month project for Carl  Heiles. Coming back to the PhD thesis, it concerned a problem of astrodynamics, the dynamics of globular clusters. In the words of Michel H\'enon (this is an extract of a CV):
" In my thesis (1961), I proposed that
the evolution of a cluster under the effect of binary encounters
leads to the formation of an infinite density cusp at the center...Later (1966), I
suggested the use of a Monte Carlo approach in order to speed up
the numerical computation of cluster evolution...This allowed me in particular to simulate
what is now called the "post-collapse evolution" of a cluster and
to show that it is characterized by a general expansion of the
cluster, fed by a flow of energy out of the central binary
star."

\begin{figure}[htbp]
\centerline{
\includegraphics[height=2.0in]{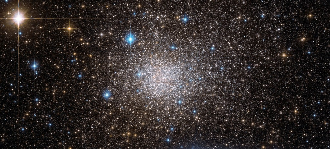}
}
\caption{Terzan $5$, from Hubble Space Telescope.
}
\label{Fig:terzan}
\end{figure}

What makes globular clusters interesting and challenging is, among other things, their high central density and their spherical shape.  They are local bound states
 of approximately ten thousand to one million stars, spread over a volume of several tens to about 200 light years in diameter. For example Terzan 5, see figure \ref{Fig:terzan}, which is located  deep within our Galaxy and slightly above the galactic plan, has the highest density of stars of all known globular clusters, about $3000$ times the average concentration of stars in the Milky Way; moreover it contains the largest number of millisecond radio pulsars (rapidly rotating binary neutron stars).

  Compared to other physical "many body" systems, globular clusters have the enormous advantage for theoreticians to depend on Newton's attraction, a very well defined interaction law. But, from the point of view of statistical mechanics, the long range character of the gravitation potential makes impossible a study by Gibbs-Boltzmann statistical mechanics and one has to turn to other methods.  This is what  H\'enon did in his PhD work. He mixed two approaches. At the time (early nineteen sixties), numerical computation was in a very early stage but H\'enon tried nevertheless to compare results of his numerical approach with those of his analytic work.

He discovered that globular clusters, seen as an initial value problem with many interacting stars, tend to form, after a finite time, what he called an "infinite density cusp" at their center (the word "cusp" is in his resum\'e, not in the original text of his thesis). This cusp is only a mathematical idealization valid in the limit of a continuous density, instead of a gas of discrete point masses. He understood as well that, once the cusp is formed, in the real physical system of discrete masses, this center is actually a place where binaries form. He considered those pairs as a sink of negative energy, so that the singularity at the center is a source of positive energy for the rest of the cluster.

In this contribution we comment on these results, after many others who have investigated after H\'{e}non's work the
theory of self-gravitating systems (see the recent papers \cite{mouhot} and \cite{PHC-2013-kyn}
 and references herein).
In the next section we comment on the fact that statistical mechanics of self-gravitating systems is an out-of-equilibrium problem, and we illustrate the role of pair formation in the simple 3-body case of Jupiter-Saturn orbiting around the Sun. In section \ref{sec:equations} we present the kinetic approach to the $N$ body self-gravitating problem, restricted to the collisionless (mean-field) case which is described by Vlasov-Newton (also named Vlasov-Poisson throughout this paper, \cite{JeansVlasov}) equation and explain how it can be solved by adding some conditions. We discuss more specifically the existence and properties of cusp-core \textit{steady-state} solutions with finite mass and energy, obtained within the frame of isotropic distribution of both velocity and position. The solution we consider have  singular  potential $\Phi(r)$ (and density $\rho(r)$) at the center of the cluster. The potential obeys a modified Lane-Emden equation (modified to ensure convergence of the total mass and energy of the cluster).
We show in subsection \ref{subsec:dim} that a one parameter family of cusp-like solutions exists besides the steady solutions, with same mass. They display an oscillating behavior around the steady cusp-core solution, the amplitude of the oscillations increasing towards the center of the cluster as $\delta \rho \sim r^{-5/2}\sin (\Omega \ln(r))$, namely periodic versus the variable $\ln(r)$. In subsection \ref{susec:nearcusp}  we comment on the way to treat the problem close to the center where a continuous Vlasov-type description is not valid. We define a region $r < r_k$ where the graininess of the density must be taken into account (where Vlasov-Poisson equation is not valid). This requires to solve the full dynamical problem of interacting point masses,  and suggests to use a matching method for connecting this solution to the outer Valsov-Poisson one.
In section \ref{sec:exact} we discuss the role of the angular momentum. We point out the irrelevance of the isotropic velocity hypothesis in a dynamical problem starting with zero angular momentum, nevertheless we show that in this case a steady cusp-core solution exists if a pure radial motion is assumed, i.e. with zero angular momentum. This conclusion leads us to reconsider in section \ref{sec:collapsedust}  the gravitational collapse of a dust gas where initially masses are immobile.  Such a dust gas has been treated by Mestel, Larson and  Penston \cite{Penston} shortly after H\'{e}non's defense. They give an explicit analytical solution for its dynamics up to  a finite time, at which the core density becomes singular . We suggest to search an after-collapse solution for this dust gas case which differs drastically from the fluid case where the matter accumulates at the center without crossing it. In the case of a dust gas, the flux of particles with negative velocity (inward motion) may cross the central part of the medium and continue their way with positive velocity (outward motion). This yields a multivalued velocity field, which is impossible in dense matter, but allowed in a dust gas. Somehow the dust gas represents a prototype for the study of evolution of a globular cluster or more generally of a system of self-gravitating point masses.

\section{Self gravitating systems as out of equilibrium systems}
\label{sec:out-of-eq}
%This point of view (formation of pairs near a singularity of density) may be understood in a slightly different way:
 Gibbs-Boltzmann equilibrium theory cannot be applied to globular clusters because there is no well defined microcanonical ensemble. For a finite number $N$ of an \textit{isolated} set of masses $m$ located in a finite domain, the thermodynamical quantities are well defined: assuming that all microstates are equiprobable (a postulate for Hamiltonian systems), one can show that the most probable distribution $f(\textbf{r},\textbf{v})$ of a macrostate ( specified by the number ${}n_i$ of particles contained in the $\nu$ microcells with size $(h/m)^3$, namely $f=n_im/\nu(h/m)^3$ ), is the one which maximizes the Boltzmann entropy

  \begin{equation}
  S= - \int \frac{f}{m}\,\, ln \frac{f}{m} \,\, d\textbf{r} d\textbf{v},
   \label{Eq:boltzentrop}
   \end{equation}

 at fixed mass $M=Nm$ and energy.  That gives the mean-field Maxwell-Boltzmann distribution
 \begin{equation}
  f(\textbf{r},\textbf{v})= A e^{-\beta m(\frac{v^2}{2}+\Phi(\textbf{r}))},
   \label{Eq:boltzdistrib}
   \end{equation}
 where the gravity potential is a solution of the Boltzmann-Poisson equation $ \Delta \Phi= 4\pi G A'e^{-\beta m\Phi}$, and where $\beta$ is the inverse temperature.  Moreover for non-rotating systems the maximum entropy state  is spherically symmetric. However the above approach leading to the global maximum of the entropy fails to describe globular clusters because the density of states
  \begin{equation}
  g(E)= \int \delta (E-H) \prod  d\textbf{r}_i d\textbf{v}_i,
  \label{Eq:dens-micro}
  \end{equation}
  diverges when some particles can go to infinity or can get a very large kinetic energy (in equation (\ref{Eq:dens-micro}) the energy is  given by the relation
%  \begin{equation}
 $ H= \frac{1}{2}\sum_{i=1,N}mv_i^2 \,\,- \sum_{i<j} G\frac{m^2}{|\textbf{r}_i-\textbf{r}_j|}$).
 %, \label{Eq:Hamilt}  \end{equation}.
 It means, among other things, that thermodynamical concepts like temperature, entropy, and so on cannot be used to describe globular clusters even with a finite number of stars (see below the 3 body case). Therefore no equilibrium state exists, and the equilibrium state (\ref{Eq:boltzdistrib}) will never be reached. However it could be possible that long-lived metastable states exist, that are local maxima of $S([f])$ at fixed $E$ and $M$. Kinetic theory is required  to inform about the time scales and to predict which metastable equilibrium state are reached from given initial data. This does not necessarily mean that globular clusters, as we see them now, are the result of the evolution of an isolated set of $N$ stars, As argued in \cite{Kaz} they could well be the result of a slow aggregation of stars already present in the Galaxy.
 The microcanonical distribution

  \begin{equation}
  P_N(\textbf{r}_i \textbf{v}_i,t)= \frac{1}{g(E)} \delta (E-H),
  \label{Eq:micro-distrib}
  \end{equation}

 is defined in general for a system of energy $E$
 as an uniform distribution of probability on the surface (or manifold) of constant energy. This probability distribution, to be well defined, requires that the total probability $g(E)$ is well defined too. This total probability is the volume of the manifold of constant energy. In the case of gravitational interaction, this manifold of constant energy has an infinite volume so that the microcanonical partition function, or density of states $g(E)$, does not exist (it is given by a diverging integral). This divergence is due to the fact that the potential energy is not bounded from below: at constant total energy, the kinetic energy of many masses may increase indefinitely at the expanse of the potential energy of a pair getting closer and closer. The most diverging part in the volume of phase space comes from configurations when only one pair becomes deeply bound to get a very large negative potential energy, although all the other stars get a large positive kinetic energy.  Usually, in a regular Gibbs-Boltzmann system with converging microcanonical partition function the point representing the state of the system explores in its dynamics the phase space uniformly in the course of time.  The wandering in phase space of finite volume of the Gibbs-Boltzmann theory is changed for systems with gravitational interaction  into an evergoing exploration of phase space. As argued in \cite{Kaz} this situation is not so unusual: one meets something similar when dealing with a point diffusing by Brownian motion in 1D: if one considers diffusion on a segment of finite length, statistical properties derived by time average are well defined and independent on time, because the equivalent of the microcanonical partition function is just the length of the segment. If one considers instead the diffusion on the full real line of infinite length (and so with a diverging microcanonical partition function), any averaging will yield a time dependent result, the average squared abscissa will grow proportional to time, etc.  Similarly, in the exploration of phase space by 3 masses (for instance) in a box  \cite{Kaz}, no average quantity exists in general, only time dependent quantities. In particular the part of phase space giving the most diverging contribution to the divergence of the microcanonical integral is the one dominating the exploration for long time. In the present problem, this maximal divergence is the one where a pair has an binding energy tending to minus infinity, an energy transferred to the kinetic energy of the other masses which grows to plus infinity, again to maximize the volume of phase space they explore.

A case of such an exploration of infinite phase space is the one of a pair of masses in the gravity field of a large immobile mass, like Jupiter and Saturn orbiting in the field of the Sun. Of course, we know well that part of the phase space for this three-body problem is filled with quasi-periodic orbits, thanks to the KAM theory. However, as the H\'enon-Heiles model shows, another part is filled by random orbits and so can be understood, at least qualitatively, by using statistical methods (instead of the perturbation methods first imagined by Newton). Therefore, in the case of this three-body problem Sun+Saturn+Jupiter it is likely that the two planets will have a chance to make a close encounter which will put enough kinetic energy into the system to allow the evasion of one planet, by transferring the kinetic energy of the two close-to-Keplerian orbits around the Sun into the kinetic energy needed for the evasion. In this respect it is well possible that the 3-body problem bears some similarities with the $N$-body problem, $N$ large, where close interaction/collisions are responsible of the evaporation of stars out of the cluster \cite{Kaz}.  One may conjecture that the KAM part of phase space (associated to quasiperiodic motion, not described by a non smooth invariant measure) will become of smaller and smaller relevance as the number of planets increases. Therefore planetary systems with more than three or four big planets will likely not survive long because of this possibity of evasion after close encounter between two planets orbiting in the field of the star.  In this respect it is relevant to point out the often observed "numerical" fact that ergodicity sets-in in non integrable systems with "many" degrees of freedom as soon as the number of freedom grows. One may conjecture that the relative weight of the KAM part of phase space decreases with $N$, number of freedom, in an exponential way, as does the chance of  falling on quasi-periodic orbits with randomly chosen initial conditions.

This example of planets orbiting in the gravitational field of a big star helps to explain what happens near the cusp of density at the center of a cluster. Close enough to this cusp, the stars cannot be considered anymore as belonging to a continuum. H\'enon derived from this remark an explanation for the pairing of stars occurring near the cusp and the idea that, because pairs may have a large negative binding energy, the cusp is a provider of positive energy to the rest of cluster. This is discussed in subsection \ref{susec:nearcusp} below.

\section{Equations and cusp-like solution}
\label{sec:equations}
In this Section we outline the analytical approach to the mean-field theory of a large number $N$ of interacting point masses. As explained in \cite{Kaz}, by taking advantage of this large number, one can carry a formal expansion in inverse powers of $N$ such that, at leading order, the dynamics is given by the solution of the Vlasov-Newton (also called Vlasov-Poisson or mean-field equation \cite{JeansVlasov}). At next order with respect to  the small parameter $1/N$ one has to add the Boltzmann collision operator as a small perturbation, etc. The fundamental mathematical object of this mean field theory is the one-body distribution function, denoted as  $f({\bf{r}}, {\bf{v}}, t)$ below, a function with positive or zero values depending on time $t$, of the position ${\bf{r}}$ and velocity ${\bf{v}}$ of a star (boldface are for vectors). Mathematically this function does not have to be smooth: it can have all sort of discontinuities, the only constraint is that it must be positive or zero, and yield finite mechanical invariants: total mass, energy and angular momentum. Moreover to make things simpler we shall assume that stars have all the same mass. We shall assume the validity of a coarse grained (or mean-field) picture, where  each star moves in the average field of the others, which is correct in the limit $N$ large if the density remains finite. The main difference with ordinary thermodynamics is that, a priori, one does not know for this system the form of this one-body distribution function in a steady equilibrium state. In a regular thermodynamic system at equilibrium, the velocity distribution is just a Maxwellian. Here this cannot be true for point masses, because this velocity distribution should be multiplied by a function of the position representing something like a Laplace equilibrium of positions in the self-consistent gravitational field. This would be in obvious contradiction with the assumption that this describes an equilibrium state because it yields a non zero weight to initial conditions of stars escaping to infinity if their velocity is large enough. Therefore there is no equilibrium one-body distribution function, and one should instead look at the steady solutions of the Vlasov-Newton equation written below.

We give first the Vlasov-Newton equation and explain how it can be solved by adding some conditions. Then we discuss more specifically the existence and properties of solutions with a singularity of density near the center of the cluster, a singularity studied first by Michel H\'enon.

The number density of stars $\rho({\bf{r}}, t)$ is normalized in such a way that
\begin{equation}
   \int\,{\mathrm{d}}^3{\bf{r}} \,\rho({\bf{r}}, t) = N \textrm{.}
   \label{Eq:nombretotal}
   \end{equation}
   The position-velocity distribution $f({\bf{r}}, {\bf{v}},
  t)$ and the number density are related by the formula:

  \begin{equation}
  \int\,{\mathrm{d}}^3{\bf{v}} \,f({\bf{r}}, {\bf{v}},
  t) = \rho({\bf{r}}, t) \textrm{.}
  \label{eq:integra}
  \end{equation}

  The Vlasov-Newton equation describes the evolution of  $f({\bf{r}}, {\bf{v}},
  t)$. As it is first order it keeps formally the property that if initially $f({\bf{r}}, {\bf{v}},
  t)$ is positive, it remains so at later (and former) times. This equation reads:
    \begin{equation}
      \frac{\partial f({\bf{r}}, {\bf{v}},
	t)}{\partial t} + {\bf{v}}\cdot\frac{\partial f({\bf{r}}, {\bf{v}},
	t)}{\partial {\bf{r}}} + {\bf{\Gamma}}({\bf{r}},t)\cdot\frac{\partial
f({\bf{r}}, {\bf{v}},
	t)}{\partial {\bf{v}}} =\;0
\textrm{.}
 \label{Eq:vlasnew}
	 \end{equation}
	 with
 \begin{equation}
	      {\bf{\Gamma}}({\bf{r}},t) = + G m \int \,{\mathrm{d}}^3{\bf{r}}'\,
	      \rho({\bf{r}}',t)
	      \,\frac{
	      ({\bf{r}}'- {\bf{r}})}{|{\bf{r}}'- {\bf{r}}|^3}
	\textrm{.}
	 \label{Eq:vlasgamma}	
  \end{equation}
  The steady solutions of Vlasov-Newton in the spherically symmetric case (no global rotation) depend on the two constants of motion in a spherically symmetric potential, the energy $E$ and the square of the angular momentum. Let assume also (as did H\'enon) that the velocity distribution is isotropic,  $\mathrm{d}^3{\bf{v}} = 4\pi v^2 \mathrm{d}v$, then the distribution depends on the energy only, and Poisson equation

   \begin{equation}
     \frac{1}{r^2} \frac{{\mathrm{d}}}{{\mathrm{d}}r}\left(r^2  \frac{{\mathrm{d}}\Phi}{\mathrm{d}r}\right)
	     \,=\,4
	     \pi G \rho(r,t)
    \textrm{,}
	 \label{Eq:poisson}
	 \end{equation}
  writes
  \begin{equation}
	      \frac{1}{r^2} \frac{{\mathrm{d}}}{{\mathrm{d}}r}\left(r^2  \frac{{\mathrm{d}}\Phi}{\mathrm{d}r}\right)
	     \,=\,16
	     \pi^2 G m  \int_{\Phi}^{0}\,{\mathrm{d}}E\,f(E) \sqrt{2(E - \Phi)}
	\textrm{.}
	 \label{Eq:equphi0}
	 \end{equation}
	 where
	
\begin{equation}
  E = \frac{1}{2} v^2 + \Phi(r)
\textrm{,}
	 \label{Eq:ener}
	 \end{equation}
  and where $\Phi$ is related to $\Gamma$ by
\begin{equation}
  {\bf{\Gamma}}({\bf{r}}) = - \frac{{\bf{r}}}{r}
  \frac{{\mathrm{d}}\Phi(r)}{{\mathrm{d}}r}
  \mathrm{.}
	 \label{Eq:Gamma}
	 \end{equation}

  This set of equations can be solved explicitly \cite{Antonov} by writing $r$ as a function of $\Phi$ and inverting Abel's transform on the right-hand side of Equation (\ref{Eq:equphi0}).
  This solution is not very easy to handle, so that we shall deal below with the equations in the  form (\ref{Eq:vlasnew}-\ref{Eq:equphi0}).

  An important quantity is the value of the gravity potential at $r = 0$. By taking $\Phi(r)$ equal to $0$ at $r$ infinity, and assuming that $ r\frac{\mathrm{d}\Phi(r)}{\mathrm{d}r}$  vanishes at $r=0$ and at infinity, the value of the potential at the center of the spherical cluster, when it exists, is equal to
    \begin{equation}
    \Phi_{0} = - \,16
				    \pi^2 G m \int_{0}^{+\infty} {\mathrm{d}}r\, r^2
				    \int_{\Phi}^{0} {\mathrm{d}}E
				    \,f(E) \sqrt{2(E -
				    \Phi)}
\textrm{.}
	 \label{Eq:phi0}
	 \end{equation}
				
 Here we shall consider another possibility,  the case where this integral diverges for
				    $r$ close to zero, making $\Phi_{0}$ infinitely negative. This
				    does not violate any basic principle, but requires that $\rho(r)$ diverges as $r$ tends to zero, but not too strongly in order that the total mass and energy of the globular cluster remains finite. This is the situation that we consider below in subsection \ref{subsec:cusp}. Before to do this, we look first at some properties of Vlasov-Newton equation.
\subsection{Properties of Vlasov-Newton equation}
\label{subsec:VNE}				
We already pointed out some properties of Vlasov-Newton kinetic equation (\ref{Eq:vlasnew}), like the fact that if its solution is positive or zero at some time it remains so at later times. Other properties are the conservation of mass (and number of stars, two non equivalent properties if stars have unequal masses, something we do not assume) of total momentum, angular momentum and total energy. Those properties are well known, and we shall only state them.
The total number of stars has been already defined in equations (\ref{Eq:nombretotal})-(\ref{eq:integra}),
the linear momentum is
  \begin{equation}
 {\bf{P}} =  m \int\,{\mathrm{d}}^3{\bf{r}}  \int\,{\mathrm{d}}^3{\bf{v}} (\bf{v}) f({\bf{r}}, {\bf{v}}, t)
  \mathrm{,}
   \label{Eq:P}
	 \end{equation}
where $m$ is the mass of each star. The angular momentum is
 \begin{equation}
 {\bf{L}} =  m \int\,{\mathrm{d}}^3{\bf{r}}  \int\,{\mathrm{d}}^3{\bf{v}} \ ({\bf{r}} \times {\bf{v}}) f({\bf{r}}, {\bf{v}}, t)
 \mathrm{,}
  \label{Eq:L}
	 \end{equation}
  where $({\bf{r}} \times {\bf{v}})$ is the usual cross product of two vectors. Lastly the total energy reads

 \begin{equation}
 {\mathcal{E}}  = m \int\,{\mathrm{d}}^3{\bf{r}}  \int\,{\mathrm{d}}^3{\bf{v}} \left(\frac{v^2}{2} + \Phi({\bf{r}}, t) \right) f({\bf{r}}, {\bf{v}}, t) \mathrm{,}
  \label{Eq:E}
	 \end{equation}
 where $ \Phi({\bf{r}}, t)$ is the gravity potential, solution in this general case of the 3D Poisson equation
  \begin{equation}
  \nabla^2 \Phi = 4 \pi G \rho({\bf{r}}, t)
  \mathrm{,}
  \label{Eq:Poisson}
	 \end{equation}
 $\rho({\bf{r}}, t)$ being related to $f(.)$ by Equation (\ref{eq:integra}). This completes the list of the mechanical invariants, mass, linear and angular momentum and energy. In the present case there is also an infinite number of other invariants, namely the values of the distribution function. This is because the distribution function obeys an equation of the Liouville type. The equation (\ref{Eq:vlasnew}) expresses that the probability density $f({\bf{r}}, {\bf{v}}, t)$ is conserved along the flow lines of the velocity field defined by the equation of  motion
 $$ \dot{\bf{r}} = {\bf{v}} \mathrm{,}$$
 and
 $$ \dot{\bf{v}} = {\bf{\Gamma}}({\bf{r}}, t) \mathrm{.}$$
 Because of Liouville theorem the 6D volume element  $({\mathrm{d}}^3{\bf{r}} \ {\mathrm{d}}^3{\bf{v}}) $ in phase space is conserved in the course of time, so that the probability density in phase space is conserved along the flow lines. Therefore any integral like
 $$<J(f)> =  \int\,{\mathrm{d}}^3{\bf{r}}  \int\,{\mathrm{d}}^3{\bf{v}} f({\bf{r}}, {\bf{v}}, t) J(f({\bf{r}}, {\bf{v}}, t))\mathrm{,}$$
 is constant in the course of time, for any function $J(f)$ such that $<J(f)>$ is given by a convergent integral. This puts various constraints on the solutions of the Vlasov-Newton equation. First, because the values of $f$ are convected along the flow lines of the equations of motion just written, an infinite value of $f$, as the one discussed below, has to be also present in the initial conditions. Moreover this pointwise conservation of the values of $f(.)$ implies that there is no restoring mechanism to ensure the smoothness of $f(.)$, this is because there is no diffusion of $f(.)$ perpendicular to the flow lines defined by the equations of motion written above. Therefore the conservation of the values of $f(.)$ during the evolution does not help much to find the steady state reached from given initial data: adding a small perturbation to this initial data with a fast variation in phase space will change the distribution of values of $f(.)$ to first order, but without any effect on the dynamics, because the acceleration  ${\bf{\Gamma}}$ is given by an integral over space which smooths out the fast variations of $f(.)$ in phase space. Because of the absence of diffusion in phase space in the direction normal to the flow lines, there is no obvious way to tackle this problem of finding the steady state resulting from a given initial condition for the Vlasov-Newton equation. Nevertheless one can predict that there is no hope to get a distribution function with a finite time singularity if the initial data for $f(.)$ are uniformly bounded because they have to remain so forever.

 There exists an exact free-fall solution of a sphere of dust ({\it{i.e}} a gas without pressure interacting only by the gravitational interaction) leading to a density cusp at the center after a finite time depending on the initial distribution of density inside the sphere, see section \ref{sec:collapsedust}. This does not contradict the property that an initially bounded solution of Vlasov-Newton remains so at any time, because this gas of dust corresponds to a velocity distribution which is initially a Dirac delta function, obviously not a smooth bounded function of the velocity.

	\subsection{Steady-state cusp-core solution}
\label{subsec:cusp}
 We consider the case $\Phi(r)$ tending to $-\infty$ as $r \rightarrow 0$,  and try to find a distribution $f(E)$ such that the solution of the spherical Vlasov-Poisson system (\ref{Eq:vlasnew}-\ref{Eq:equphi0}) has finite mass and energy. We use the most straightforward method, starting with an assumed form of $f(E)$, integrate over the velocity  to obtain the density $\rho$ , equation (\ref{eq:integra}), and solve Poisson equation (\ref{Eq:equphi0}) , to get the corresponding potential $\Phi$.
 Let us recall the behavior of the solution if the energy distribution is given by  the most simple formula, the truncated power law
  \begin{equation}
  f(E)= k_f(E_0-E)^{n-\frac{3}{2}}Y(E_0-E)
  \textrm{.}
	 \label{Eq:fE}
	 \end{equation}
 with $n$ real positive exponent, $E_0$ (maximum energy) is a constant which is introduced to ensure that $f(.)$ has compact support, $Y(.)$ is Heaviside function and $k_f$ a normalizing constant. In the following the distribution $f(.)$ has no compact support because we shall assume that the energy can get infinite negative values. The mass-density is computed via equation (\ref{eq:integra}), integrating by parts over the velocities and setting $v^2= 2 (E_0-\Phi) cos^2\theta$ we obtain $ \rho=2^{\frac{3}{2}}\pi B(\frac{3}{2},n-\frac{1}{2})k_f (E_0-\Phi)^n$ where $B(a,b)$ is the Beta function, or

 \begin{equation}
  \rho(r)=c_n ( E_0-\Phi)^n
  \textrm{,}
	 \label{Eq:rho}
	 \end{equation}
with  $c_n= (2\pi)^{3/2} \frac{(n-3/2)!}{n!}k_f$.

Assuming that the maximum energy is negative, $E_0 <0$,
equation (\ref{Eq:equphi0}) can be written in a scaled form as
				
\begin{equation}
	     \frac{1}{\tilde{r}^2} \frac{{\mathrm{d}}}{{\mathrm{d}}\tilde{r}}\left(\tilde{r}^2  \frac{{\mathrm{d}}\tilde{\Phi}}{\mathrm{d}\tilde{r}}\right)
	     \,= (- 1 - \tilde{\Phi})^{n} Y(- 1 - \tilde{\Phi})
	\textrm{.}
	 \label{Eq:equphi0resc}
	 \end{equation}
when using the scaled variables $\tilde{\Phi}=\frac{\Phi}{-E_0}$, $\tilde{r}= \frac{r}{b}$  with $ b=[4 \pi G c_{n}(-E_0)^{n-1}]^{-1/2}$.
%We have to add the condition that $\Phi \to 0$ as $r \to \infty$.
Finally defining the quantity  $ \psi= -1-\tilde{\Phi}$, which is positive inside the cluster and tends asymptotically to $-1$ as $r \to \infty$, we obtain the Lane-Emden equation for the "relative" (with respect to $(-1)$) potential $\psi$,

\begin{equation}
	    - \frac{1}{\tilde{r}^2} \frac{{\mathrm{d}}}{{\mathrm{d}}\tilde{r}}\left(\tilde{r}^2  \frac{{\mathrm{d}}\psi}{\mathrm{d}\tilde{r}}\right)
	     \,= \psi^{n} \,Y(\psi)
	\textrm{.}
	 \label{Eq:lane-emden}
	 \end{equation}
The literature on this equation usually rejects solutions with a density cusp at $r =0$ because it leads to infinite mass when considering its behavior at large $r$. For that reason we shall consider separately the behavior of the power law solution of  Lane-Emden equation as $r$  tends to zero (where $\psi$ tends to $+\infty$)   and $r \gg 1$ where $\psi$ goes to zero.

Let us first consider
the behavior of a diverging power law solution for $\tilde{r} \ll 1 $, and find the range of parameter $n$ which gives finite mass and energy in the central part of the  sphere. At large negative $E$ , the relation $f(E)\simeq k_f(-E)^{n-\frac{3}{2}}$ inserted into  Poisson's equation yields
				
 \begin{equation}
\tilde{\Phi}_n\simeq -A_n \tilde{r}^{\left(\frac{-2}{n-1}\right)}
\textrm{,}
	 \label{Eq:Phin}
	 \end{equation}
or
 \begin{equation}
\psi \simeq A_n \tilde{r}^{\mu}
\textrm{,}
	 \label{Eq:psin}
	 \end{equation}

where the negative exponent
 \begin{equation}
\mu = \frac{-2}{n-1}
	 \label{Eq:mu}
	 \end{equation}
 is consistent with the behavior of
				    $f$ at large negative $E$.  The expansion of Poisson equation close to $r=0$ gives
 \begin{equation}
A_n= \left(\frac{2(n-3)}{(n-1)^2}\right)^{\frac{1}{n-1}}
\textrm{.}
	 \label{Eq:An}
	 \end{equation}
				
\begin{figure}[htbp]
\centerline{
(a)\includegraphics[height=1.5in]{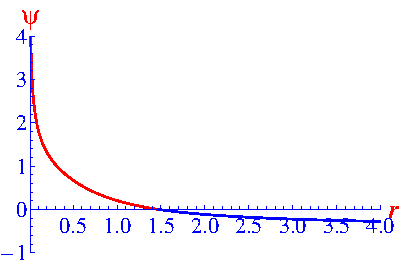}
(b)\includegraphics[height=1.5in]{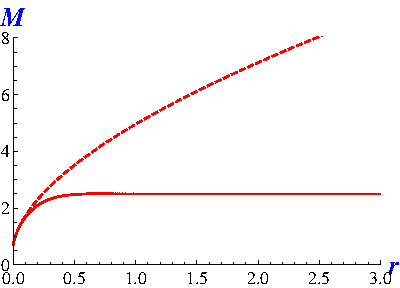}
(c)\includegraphics[height=1.5in]{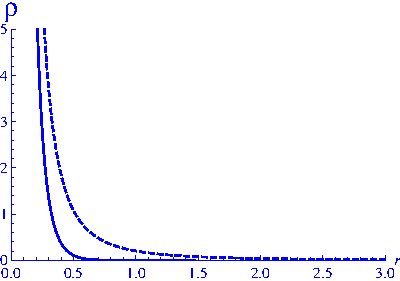}
}
\caption{Cusp-core solution of the modified Lane-Emden equation for $ n=6$ and $n'=2$, $k_f=k_{f'}$, $E_0=-1$ or $k_{n n'}\simeq 6.1$. (a) Relative potential $\psi(\tilde{r})=-1-\tilde{\Phi}(\tilde{r})$, (b) Mass enclosed inside a sphere of radius $\tilde{r}$, (c) density $\rho(\tilde{r})$. The solid (resp. dashed) lines are solutions of the modified Lane-Emden equation with $n=6$ and $n'=2$, (resp.  of the "usual" Lane-Emden equation with $n=6$, $n'=0$).
}
\label{Fig:phi-6}
\end{figure}

Close to $r=0$ the total energy converges if the integral $\int_{0}^{\infty}{\mathrm{d}}rr^2\int_{\phi}^{0}
{\mathrm{d}}E E f(E)\sqrt{2(E-\Phi)}$ converges. This requires

\begin{equation}				
n> 5\mathrm{,}
\label{Eq:condn}
\end{equation}

that is equivalent to the condition $\mu > -1/2$ for the negative exponent in $$\Phi(r)|_{r \to 0} \simeq -A'_n r^{\mu} \mathrm{.}$$
Note that the condition of convergence of mass is less stringent. The critical value of the exponent $n=5$ is exactly the classical critical power beyond which
%no physical "smooth" solution (with finite value at $r=0$) of
the " smooth" (with finite central density) solution of Lane-Emden  equation has infinite radius and mass. This divergence of radius and mass also occurs for the cusp-like solution with $n$ larger than $5$, as illustrated in fig.(\ref{Fig:phi-6})-b, dashed line.

 Let us consider the outer part of the cluster, the domain $\tilde{r} >>1$. In this domain, the relative potential must tend to zero. Assuming a power-law behavior for the energy distribution, $f(E)\simeq k'_f (E_0-E)^{n'}$, and considering the case $E\rightarrow E_0$, it can be easily shown that the condition for the mass to be finite is
  \begin{equation}
  n' < 3
  \textrm{,}
	 \label{Eq:nprime}
	 \end{equation}
 the condition on the energy being less drastic.
  Finally, any expression of $f(.)$ which behaves at small $r$ like $ f(E) \simeq(-E)^{n-\frac{3}{2}}$, and at large $r$ as $ (E_0-E)^{n'-\frac{3}{2}}$ with $n$ and $n'$ fulfilling respectively the relations (\ref{Eq:condn}) and (\ref{Eq:nprime}), could be physically relevant with a cusp at the center.

  As an example the function $f(E)= k_f(E_0-E)^{n'-\frac{3}{2}}[1+ k (E_0-E)^2]^{\frac{n-n'}{2}}$ could be a candidate, or
   the simplest form,
 \begin{equation}
  f(E)= k_f(E_0-E)^{n-\frac{3}{2}} + k'_f(E_0-E)^{n'-\frac{3}{2}}
 \textrm{,}
\label{Eq:fsum}
\end{equation}

that we shall investigate because it is analytically tractable. Inserting expression (\ref{Eq:fsum}) for $f(E)$ in equation (\ref{eq:integra}), the density becomes the sum of two terms,

 \begin{equation}
  \rho(r)= c_n (E_0-\Phi)^n +c_{n'}(E_0-\Phi)^{n'}
  \textrm{,}
	 \label{Eq:rhosum}
	 \end{equation}
 so that using the scaled variables defined just above, the equation for the relative potential $\psi(r)$ writes

 \begin{equation}
	    - \frac{1}{\tilde{r}^2} \frac{{\mathrm{d}}}{{\mathrm{d}}\tilde{r}}\left(\tilde{r}^2  \frac{{\mathrm{d}}\psi}{\mathrm{d}\tilde{r}}\right)
	     \,= \left(\psi^{n} + k_{n n'} \psi^{n'} \right) \, Y(\psi)
	\textrm{,}
	 \label{Eq:2lane-emden}
	 \end{equation}
where $k_{n n'}=\frac{c_{n'} (-E_0)^{n'}}{c_{n} (-E_0)^{n}} $.  Defining a scaled density $\rho$ by dividing the density (\ref{Eq:rhosum})  by the constant $c_n(-E_0)^{n}$, we obtain (keeping the same notation)
\begin{equation}
\rho(\tilde{r})=\psi^{n}(\tilde{r}) + k_{n n'} \psi^{n'}(\tilde{r})
\textrm{,}
	 \label{Eq:density}
	 \end{equation}
 a positive quantity inside the cluster, equal to zero outside.
The solution of this modified Lane-Emden equation is shown in figure (\ref{Fig:phi-6})  for the values of the exponents $n=6$ and $n'=2$ that satisfy the condition for finite mass and energy. The relative potential in (a) is drawn in red inside the globular cluster, vanishes at $\tilde{r}=r_g$  and tends asymptotically to the value $-1$  ($\Phi$ tends to zero) as $\tilde{r}$ goes to infinity (blue part of the curve). This blue part of the curve corresponds to the outside of the globular cluster where the relative potential $\psi$ fulfills equation (\ref{Eq:2lane-emden}) with zero in the r.h.s.
and behaves asymptotically as $1/\tilde{r}$. The mass $M(r)$ enclosed inside a sphere of radius $\tilde{r}$,
 \begin{equation}
M(\tilde{r})=4\pi \int_0^{\tilde{r}} \mathrm{d}r' r'^2 \rho(r')
\textrm{.}
	 \label{Eq:masseM}
	 \end{equation}
is drawn
in (b), and the density $\rho(r)\,Y(r-r_g)$ given by equation (\ref{Eq:rhosum}) is drawn in (c) (both with solid lines). The standard Lane-Emden case (with $n=6$ ) is shown in dashed lines for comparison.

%The literature on Lane-Emden equation usually rejects solutions with a density cusp at $r =0$.
%For the power law exponent on the right-hand side of this equation, this excludes that this right-hand side is a function of $\Phi$ which does not grow faster than $(-\Phi)^5$ as $\Phi$ tends to minus infinity.
Let discuss the physical relevance of such solutions presenting a density cusp at $r=0$.
 In the interior of a star a diverging density could be excluded on physical grounds. However, if $\Phi$ represents the gravity potential there, it is clear that this divergence of a non relativistic gravity potential should be stopped by general relativity effects: such effects depend on the ratio $|\Phi|/c^2$, $c$ speed of light, and $\Phi$ gravity potential per unit mass, Therefore, in very dense matter near the core of a star, the indefinite growth of $(-\Phi)$ as $r \to 0$ in the classical limit could only mean that there is  a sphere near the core where general relativity effects have to be taken into account. In this respect the case of steady state of globular clusters with $n >5$  is a priori different of the core of a dense star: near the center of the globular cluster, one should instead take into account the graininess of the mass density, namely replace the continuum described by Vlasov-Newton by a discrete set of interacting masses as explained below, subsection \ref{{susec:nearcusp}}. In principle one could also have situations in globular clusters where the regularization of the divergence of $\Phi(r)$ near the cusp is by general relativity effects. However, the numbers one can put on the ratio $|\Phi|/c^2$ at the core of globular clusters are far too small to make general relativity effects relevant. It could be however that they become so near the center of Galaxies, but this is another story.

	\subsection{ Multiplicity of cusp-like solutions}
%Dimensionality of the stable manifold}
\label{subsec:dim}

An important question is the dimensionality of the manifold of the cusp-core steady solutions of equation (\ref{Eq:equphi0}). In other terms does it exist a family of solutions which behave as the steady cusp solution  described above?
Let $r^*$ be  any  finite value of the radius. At this point  a small deviation deviation $\delta \psi(r^*)$ of the potential has two free parameters,  the values of $\delta \psi(r^*)$ and $\delta \psi_{,r}(r^*)$. We ask if these two parameters are enough
 to reach a solution having same behavior close to $r=0$ and same total mass as the steady solution. The answer will be given
 by linearizing the solution of equation (\ref{Eq:equphi0}), or equivalently of its scaled forms (\ref{Eq:2lane-emden}).

 \begin{figure}[htbp]
\centerline{
(a)\includegraphics[height=1.3in]{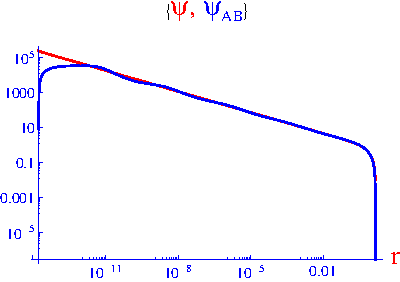}
(b)\includegraphics[height=1.3in]{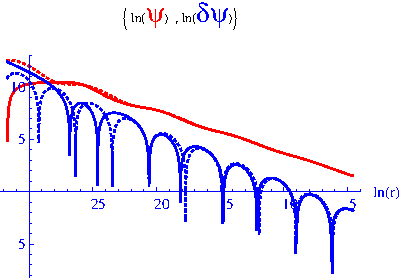}
(c)\includegraphics[height=1.3in]{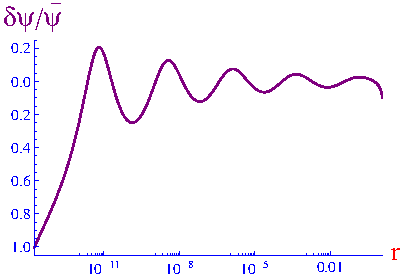}
}
\caption{Numerical solution: potential of the cusp  solutions with and without deviation.
(a) Steady ($\overline{\psi}$, red curve) and perturbed ($\psi_{AB} (\tilde{r})$, blue curve) solutions of equation (\ref{Eq:2lane-emden}).
(b) Perturbed solution $\psi(\tilde{r})=\psi_{AB}$  and deviation $\delta \psi(\tilde{r})$, solid red and blue curves respectively, compared with the first order analytical expressions (\ref{Eq:psianal})-(\ref{Eq:deltaphi}), dotted lines.
 (c) Relative deviation $(\psi-\overline{\psi})/\overline{\psi}$ versus $\ln(\tilde{r})$.
 %(c) Density $\rho_{AB}(\tilde{r})$ and $\overline{\rho}(\tilde{r})$.
 The modified Lane-Emden equation with $n=6$ and $n'=2$, is solved with i.c. $\psi(r^*)=\psi_a(r^*)$ and $\psi_{,\tilde{r}}(r^*)=(\psi_a)_{,\tilde{r}}(r^*)$ ($A=B/2= -0.035$,  $r^*=10^{-2}$).
 The dotted curves display the analytical expression (\ref{Eq:psianal}).
}
\label{Fig:family}
\end{figure}

 Let us  explore the possible behaviors of small variations of $\psi$ near the solution $\overline{\psi}(\tilde{r})$   of the modified Lane-Emden equation (\ref{Eq:2lane-emden}).
 %, which describes the cusp behavior close to $r=0$ .
 We introduce a small function $\delta \psi(r)$ such that the exact solution of equation (\ref{Eq:2lane-emden}) is $\overline{\psi}(\tilde{r}) + \delta \psi(\tilde{r})$, or equivalently $ \Phi=\overline{\Phi}(r) + \delta \Phi(r)$ with $\delta \Phi(r)=  \frac{\delta \psi(r)}{E_0}$ .

\begin{figure}[htbp]
\centerline{
(a)\includegraphics[height=1.3in]{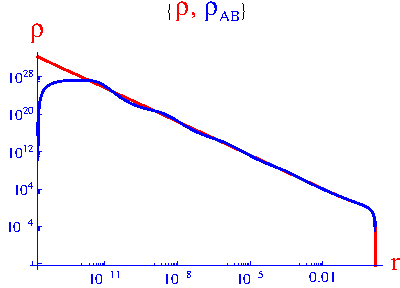}
(b)\includegraphics[height=1.3in]{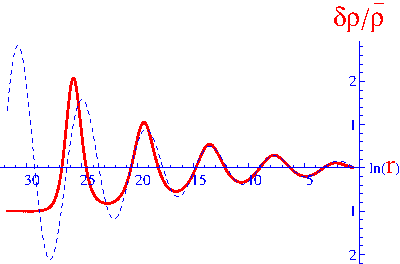}
}
\caption{Numerical solution: density of the cusp solution with and without deviation, same parameters and i.c. as in Fig.\ref{Fig:family}.
(a) Steady ($\overline{\rho}$, red curve) and perturbed ($\rho_{AB} (\tilde{r})$, blue curve) solutions of equation (\ref{Eq:2lane-emden}).
 (b) Relative deviation $(\rho-\overline{\rho})/\overline{\rho}$ versus $\ln(\tilde{r})$, the solid curve corresponds to the numerical solution, the dotted curves to the analytical expression (\ref{Eq:psianal}).
}
\label{Fig:deviation}
\end{figure}

 In the core region the potential is given by the asymptotic expressions (\ref{Eq:Phin})-(\ref{Eq:An}), and the positive relative potential is $\psi_n \simeq A_n \tilde{r}^{\left(\frac{-2}{n-1}\right)}$. The small deviation $\delta \psi$ is a solution of the linear homogeneous equation
 \begin{equation}
	     -\frac{1}{\tilde{r}^2} \frac{{\mathrm{d}}}{{\mathrm{d}}\tilde{r}}\left(\tilde{r}^2  \frac{{\mathrm{d}}\delta \psi}{\mathrm{d}\tilde{r}}\right)
	     \,= n(\overline{\psi}(\tilde{r}))^{n-1} \delta \psi \approx  k_{\delta} \tilde{r}^{- 2} \delta \psi
	\textrm{.}
	 \label{Eq:equphi1}
	 \end{equation}
	with $ k_{\delta} =  \frac{2(n-3)n}{(n-1)^2}$,  a positive constant in the domain $n> 5$ considered here.	
 The  solution of this equation, of the form $\delta \psi (r)=Ar^{-\lambda}$  is such that $\lambda $ is a root of the second degree polynomial
 \begin{equation}
 \lambda^2 -\lambda +k_{\delta}=0
 \textrm{,}
	 \label{Eq:lambda}
	 \end{equation}
 that gives
 $$ \lambda =\frac{1}{2}\pm i \Omega$$

 with $ \Omega^2=  k_{\delta} -1/4= \frac{7n^2-22n-1}{4(n-1)^2}$  which is positive  for $n > \frac{11+\sqrt{7}}{7}$. Therefore the behavior of $\delta \psi(\tilde{r})$ as $\tilde{r} \to 0$ is given by
\begin{equation}	
 \delta\psi (\tilde{r}) =  r^{-1/2} \left( A \cos (\Omega \ln(\tilde{r})) + B \sin(\Omega \ln(\tilde{r})) \right)
 \mathrm{,}
	 \label{Eq:deltaphi}
	 \end{equation}
 where $A$ and $B$ are arbitrary constants. Note that equation (\ref{Eq:equphi1}) is homogeneous with respect to the variable $\tilde{r}$. Therefore it can be transformed into a linear autonomous and homogeneous second order differential equation by changing $\tilde{r}$ into $x=\ln(\tilde{r})$, that leads to a solution of the form $e^{\lambda x}$, as expressed in (\ref{Eq:deltaphi}).  As $\tilde{r}$ tends to zero, or $x$ tends to minus infinity, the  deviation  $\delta \psi$  oscillates  periodically with respect to the variable $ln(\tilde{r})$, with a frequency $\Omega$  of order unity, and increasing amplitude.
  We have to notice that the periodic behavior of $\delta \psi$ and $\delta \rho$ in terms of $\ln(\tilde{r})$,  see Figs.\ref{Fig:family}-\ref{Fig:deviation}, corresponds to an aperiodic oscillatory behavior of the solution with respect to the physical variable $\tilde{r}$, with shorter and shorter period as the radius $r$ tends to zero. This ensures a finite mass  in the cusp region, where the integral of the density (multiplied by $\tilde{r}^2$) is practically insensitive to the more and more rapidly oscillating perturbation, as confirmed by the numerical study.

The numerical study of the family of perturbed solutions is performed by taking initial conditions in the cusp region, at a given radius $\tilde{r}=r^* <<1$ where equation (\ref{Eq:2lane-emden}) reduces to (\ref{Eq:lane-emden}). There the solution writes at first order with respect to a small deviation $\delta \psi$, as
  \begin{equation}	
\psi(r^*) = A_n (r^*)^{\mu}+ \delta \psi( r^*)
 \mathrm{,}
	 \label{Eq:psianal}
	 \end{equation}
where $\delta \psi$ is given by the relation (\ref{Eq:deltaphi}).  Equation (\ref{Eq:psianal}) is a good approximation of the perturbed solution as soon as the relation $\delta \psi(r^*)/ \overline{\psi}(r^*)<<1$ is fulfilled, as illustrated in Fig. \ref{Fig:family}-(b) where the dotted curves stands for the analytical expression (\ref{Eq:deltaphi}). In Figs. \ref{Fig:family}-\ref{Fig:deviation} we took initial conditions at $r^*=10^{-2}$, where $(\delta \psi / \psi)(r^*)= -0.035$.
In fig.\ref{Fig:family}-(a) the steady solution $\psi$ (red curve) displays a straight portion with slope $\mu$ (in $\log/\log$ plot) which delimitates the cusp domain; the perturbed solution $\psi_{AB}$ (blue curve) oscillates around the steady solutions with increasing amplitude as  $\tilde{r}$ approach the center, and decreasing amplitude towards the edge of the cluster, see Fig.(c) in Log/linear scale. At a given radius $r_{b}$ much smaller than $r^*$ ($r_b \sim 10^{-14}$ for the data of the figure) the perturbed solution crosses zero, because at this radius the modulus of the negative deviation becomes as large as the steady solution value. Inside the very small domain $\tilde{r} < r_b$ the physical solution has to be taken as $\psi(\tilde{r})=0$ because of the positivity of the relative potential, that defines a sort of bubble inside which the density of stars is strictly zero, whereas it is very large outside but close to the bubble surface.  The potential $\psi$ and the modulus of the deviation $|\delta \psi |$ are drawn in $\log$ scale in
Fig.(b), together with the analytical expression (\ref{Eq:psianal}), in order to  illustrate the relative values of the oscillating potential and its deviation, and the validity domain of the first order solution.

The density $\rho(\tilde{r})$, plotted in Fig.\ref{Fig:deviation} with same notations, also displays oscillating behavior around the steady cusp-solution, with relative amplitude decreasing from the  bubble surface towards the edge of the cluster, and $\rho(\tilde{r}) =0$ inside the bubble.
In the domain $\tilde{r} > r^*$, the numerical solution clearly shows a small amplitude oscillation around the steady solution.  The constraint of mass conservation is satisfied by fitting one of the two parameters, either $A$ or $B$.
In summary the modified Lane-Emden equation with initial condition of the form (\ref{Eq:psianal}) taken at $r^* <<1$, with $\delta \psi( r^*)/  \overline{\psi}( r^*)$ much smaller than unity, has  a one-parameter family of  solutions oscillating around the steady cusp-solution, and vanishing abruptly inside a very small bubble located at the center of the cluster.

\subsection{What happens physically near the cusp?}
\label{susec:nearcusp}
%This example of planets orbiting in the gravitational field of a big star helps to explain what happens near the cusp of density at the center of a cluster.
We try now to connect the solution with a diverging density at $r=0$ with the point made by H\'enon on the cusp as a source of binaries and a source of energy.  A diverging density does point to the fact that the assumption behind the validity of the mean-field approach breaks down near the center. Below we give a more quantitative assessment of this point and explain how to define a local approximation near $r = 0$, different from the mean-field theory and how to match it with this mean-field theory sufficiently far from the center.

Close enough to the cusp, the stars cannot be considered anymore as belonging to a continuum. H\'enon derived from this remark a picture of the long term dynamics of globular clusters.  He explained first how pairs (or binaries) form in a globular cluster, supposed to be made initially of single stars. Actually, if stars are on average at large distances from each other, the chance they form a pair is very small, because this would require a very unlikely three stars encounter. Instead near the cusp there is a much larger probability to make pairs because of the large inhomogeneity of the mean-field there. Compared to the problem just alluded to, the region of the cusp is a bit like the attracting mass of the Sun in a planetary system. From equations (\ref{Eq:psin})-(\ref{Eq:mu}), the density near the cusp behaves like $\rho(r)=\sim r^{\mu n} \sim r^{\mu - 2}$. It diverges at $r = 0$ because $\mu$ is negative. To put it in a way where the physical quantities appear, we write this density like
$$ \rho(r) \approx \frac{N}{R^3} \left(\frac{r}{R}\right)^{\mu -2} \mathrm{.}$$
where $N$ is the total number of stars in the cluster and $R$ is the radius of the cluster. This writing of $\rho(r)$ is such that, when $r$ grows big enough to become of order $R$, the density becomes of order $N R ^{-3}$, the average density in the cluster up to a multiplicative constant of order one. Therefore the graininess of the density near the cusp will be significant when the radius $r$ will be such that the total number of stars between $r = 0$ and $r = r_k$ will be of order one, say a finite number $K$. This number is related to $r_k$ in such a way that $K = 4\pi \int_0^{r_k} {\mathrm{d}}r r^2 \rho(r)$. Up to an unknown multiplicative constant of order one, this yields the following relationship between $r_k$, $R$, $K$ and $N$:
$$ r_k = \left(\frac{K}{N}\right)^{\frac{1}{\mu + 1}} R   \mathrm{.}$$ Note the exponent  $\frac{1}{\mu + 1}$ is positive because $\mu$ is between $0$ and $(-1/2)$.

For $K$ very large one should recover a continuous medium, that is a medium described by the Vlasov-Newton equation. It means that, in the range $r \gg r_1$, with  $ r_1 = \left(\frac{1}{N}\right)^{\frac{1}{\mu + 1}} R $, the Vlasov-Newton equation applies and so provides the approximation for the dynamics far from the cusp. As usual in matching problems, this "asymptotic" picture is valid in an intermediate range $r_k \ll r \ll R$. In this range the power law found before applies and it describes the local solution of the Vlasov-Newton equation. At short distance instead, that is for  $ r \sim r_k$, one must solve the full dynamical problem of interacting point masses, with the condition that it matches at $r \gg r_k$ the solution of the Vlasov-Newton equation. This problem of solving at short distance with the full "molecular" dynamics (the inner problem) and matching it with the mean-field kinetic theory far away (the outer problem) is obviously rather challenging, but seemingly solvable in principle. It amounts to solve a $K$-body problem in the sphere of radius $ r_k$, with the condition that for $ r =  r_k$ masses coming with a positive radial velocity escape, although masses with a negative radial velocity are injected randomly on the surface of this sphere with a probability distribution given by $f(E)$ computed for  $ r = r_k$ by solving the Vlasov-Newton equation. In principle, if $K$ is big enough there should be a smooth interpolation between the statistics of particles inside the sphere and the continous distribution outside of it. That it is necessary to solve the full $K$-body problem inside the sphere of radius $r_k$ is a consequence of the fact that the mean-field theory requires $N$ to be very large. Inside the sphere of radius $r_k$, because of the boundary condition on the surface, the order of magnitude of the potential and of the gravitational energy of a given star are the same, as everywhere in the cluster, but it happens that $K$, number of stars, is not large, and of order one instead. Therefore, in this sphere one has to solve the full $K$-body problem. The power law behaviors derived above for the gravity potential will be imposed as asymptotic conditions for the inner problem.  The creation of pairs will occur as a small perturbation to the velocity distribution (pairs should contribute negligibly to the mass density and so to Poisson's potential). Such a creation of a population of binaries will perturb slowly the cluster near the cusp. It will require in the long run to assume that this population of binaries to be added to the population of single stars, Later on such binaries will return near the cusp and have a chance either to yield multiple stars or to be transformed again into two single stars and so on.

 \section{On the formation of a density cusp}
 \label{sec:exact}

 In the previous sections, we did analyze the existence of a cusp of density at the center of a cluster in the mean-field limit. This analysis was done for a steady state. In H\'enon original work, this cusp was considered instead as the result of a dynamical process. We look at this possibility below, namely we try to find conditions such that the initial density is everywhere finite and tends to infinity at later time at $r =0$. The occurrence of singularities in density under the effect of gravitational attraction has attracted much attention over the years. Likely the most ancient model proposed by Mestel was the one of a sphere of dust filled uniformly at time zero and collapsing under the effect of self-attraction of the dust gas 
 initially at rest \cite{Penston}.  In particular an exact solution is known in a parametric form for this \textit{pressureless} case, with various exponents for the singularity, see section \ref{sec:collapsedust}.

On the other hand pressure has been added to this picture, namely a relation between pressure and density, that leads to a description largely used to treat the collapse of massive stars, neutron stars and white dwarfs seen as compressible unviscid fluids. When this is done, no exact general solution is known. Nevertheless for a power law relation between pressure and density (polytropic fluid case), it is relatively easy to write equations  for a self-similar collapse of the first kind in Zel'dovich \cite{Zel} classification: combining the fluid equations and the power law for the pressure-density relation, one finds an unique set of exponents for the self-similar equations. These exponents are such that any physical quantity depends on time $\tau = t_c - t$, $t_c$ time where the collapse occurs, and on $r$, radius through a power of $\tau$ times a function of a monomial $\zeta = r \tau^{\beta}$. The exponent $\beta$ is found by imposing that the equations so derived are ordinary differential equations of functions of $\zeta$ although the equations one starts from are partial-differential equations for functions of $r$ and $t$, like the density and velocity field. However things are not so easy. Fundamentally, this is because even a finite time singularity does not guarantee that the self-similar solution is of the first kind. Actually, it happens in this problem that the solution found numerically \cite{Mannev} follows the laws of the self-similarity of the second kind. The pressure term, assumed to be of the same order of magnitude as the gravitational term in the first kind, becomes actually negligible compared to gravity. That transforms the problem into a second kind and brings a freedom in the exponents. The free exponent is found  from asymptotic matching arguments, because the inner self-similar solution must match the outer one, a constraint not satisfied by the first kind solution. This remark shows how careful one must be when dealing  with  singularities due to the collapse of a self-gravitating gas with pressure-like effects. In addition we have to note that the kinetic theory introduces in the dynamics something more complex than just a density-temperature dependent pressure.

\subsection{How to reach a steady distribution ?}
\label{subsec:dyncusp}
The first difficulty met when trying to extend the theory of steady states to time dependent systems concerns the validity of the isotropic velocity distribution hypothesis. A priori the distribution $f({\bf{r}}, {\bf{v}}, t)= f(E,L,t)$  depends on the energy $E$ and angular momentum modulus $L$ with $L^2=|\textbf{r}\wedge \textbf{v}|^2$ (Jeans theorem, see \cite{mouhot} for instance), then the isotropic spherical models require  that the distribution $f(E,L,t)$ is uniform with respect to the angular momentum $L$. However this assumption is disputable, and rarely fulfilled in numerical studies which generally assume zero velocity everywhere in the star, that implies a distribution depending both of the energy and angular momentum as discussed below.

\subsubsection{L=0}
Let us take an initial condition with radially symmetric distribution of position, and zero velocity, $f(\textbf{r},\textbf{v},0)= f(r)\delta (\textbf{v})$, with $\delta (.)$ the Dirac distribution and $r=|\textbf{r}|$. Then the angular momentum is null at time $t=0$, and one may write $f(\textbf{r},\textbf{v},0)=f(E) \delta (L)$.
 After an infinitesimal time, all masses will begin to fall toward the center making the velocity distribution obviously anisotropic. We have still $L=0$ since the velocity is purely radial everywhere. The inward motion is described by a distribution of the form  $f(\textbf{r},\textbf{v},t)= f(r,u)\delta (\textbf{w})$ (with $u$ the radial component of the velocity, and $\textbf{w}$ its 2D orthogonal component), which leads to the form
  \begin{equation}
 f(\textbf{r},\textbf{v},t)= f(E) \delta(L)
   \textrm{.}
	 \label{Eq:f-L0}
	 \end{equation}
Now let us see if a steady distribution with $L=0$ formally exists.  Assuming that the steady motion keeps this purely radial velocity dependance, and that close to the center the energy distribution is  given by the simple formula $f(E)= (E_0-E)^{n-\frac{3}{2}}$, the density   %$\rho(r,t)=\int_{\Phi}^{0}\,{\mathrm{d}}E\,(E_0-E)^{n-\frac{3}{2}}\delta (v_r-u)\delta^2 (\textbf{w}) $
 becomes
  \begin{equation}
 \rho(r,t)= 2\int_{0}^{\infty}(\psi-\frac{u^2}{2})^{n-\frac{3}{2}}\,{\mathrm{d}}u
  \textrm{,}
	 \label{Eq:dens-L0}
	 \end{equation}
when we insert the differential expression $d\textbf{v}= \pm du \, d^2\textbf{w}$. Finally if a pure radial motion persists until an equilibrium state is found, the steady density will take the same form as in the isotropic velocity case
%up to a multiplicative constant factor ($d_n$ in place of $c_n$) and a different exponent ($n-1$ in place of $n$),
  \begin{equation}
  \rho(r,t)= d_n(E_0-E)^{n-1}
  \textrm{,}
	 \label{Eq:dens-inward}
	 \end{equation}
with $d_n=2\sqrt{2}\int_0^{\frac{\pi}{2}}{\sin^{2(n-1)}\theta d\theta}$ .
Defining $\psi=-1-\tilde{\Phi}$ and using the same scaling as above except $b=[G d_n(-E_0)^{n-2}]^{-\frac{1}{2}}$ , Poisson equation (\ref{Eq:equphi0resc}) becomes identical to (\ref{Eq:lane-emden}) except $n \rightarrow (n-1)$ in the r.h.s. Using the results of section \ref{subsec:cusp} , we can conclude that the modified distribution (\ref{Eq:fsum}) together with the expression (\ref{Eq:f-L0}) could describe an\textit{ anisotropic} cusp-core steady solution  (with zero angular momentum) with finite mass and energy under the conditions $n>6$ and $n' <4$.

\subsubsection{ $L \neq 0$}

 Generally speaking the velocity distribution of a spherically symmetric system of masses depends on three scalars, $v^2$, $r^2$ and ${\bf{r}} \cdot {\bf{v}}$.  However the number of scalars can be reduced to two. This is because one can reduce, following Newton (Principia, Book 1, prop. 41), the equation of motion to an equation for the radius with a modified potential,
 \begin{equation}
   \ddot{r} = - \frac{G M(t, r)}{r^2} + \frac{\textit{l}}{r^3} \mathrm{,}
   \label{eq:newt}
   \end{equation}
  where $M(r)$ is the mass enclosed inside the sphere of radius $r$ defined in (\ref{Eq:masseM}) and $\textit{l}=L^2= (\frac{| \textbf{L} |}{m})^2 $ is the square of conserved angular momentum of a star divided by its mass $m$, related to the velocity by $$L^ 2 = r^2 v^2 - ({\bf{r}} \cdot {\bf{v}})^2=r^2(v^2-u^2),$$  with $u=\dot{r}$ the radial velocity.
  For a given particle the only quantities changing in the course of time are the radial position $r$ and velocity $u$, so that one can reduce the Vlasov-Newton equation for a spherically symmetric system to
 \begin{equation}
      \frac{\partial f(r, u, \textit{l}, t)}{\partial t} + u \frac{\partial  f(r, u, \textit{l}, t)}{\partial r} + \Gamma_r(r, \textit{l}, t)  \frac{\partial f(r, u, \textit{l}, t)}{\partial u} =\;0
\textrm{.}
 \label{Eq:vlasnewspher}
	 \end{equation}
where $ \Gamma_r(r, \textit{l}, t) =  - \frac{G M(t, r)}{r^2} + \frac{\textit{l}}{r^3}$ is the radial acceleration. The difference between this equation and the more general one written in (\ref{Eq:vlasnew}) is that $u$, $\textit{l}$ and $r$ are scalars, which reduces considerably the number of scalar variables. Contrary to the case of the Vlasov-Newton equation written before, the angular momentum appears explicitly in the equation, as it enters in $\Gamma_r$.  Therefore it is not possible in general to get rid of this angular momentum, because, even if the initial condition $f(r, u, \textit{l}, t =0)$ is independent on $\textit{l}$, the solution at later times is not because $\textit{l}$ appears explicitly in the equation of evolution of $f(.)$.

To make the Vlasov-Newton equation (\ref{Eq:vlasnewspher}) fully explicit, one has to write $\Gamma_r(r, \textit{l}, t)$ as a function of $f(r, u, \textit{l}, t)$. One has first the integral expression of $\Gamma_r(r, \textit{l}, t)$ as a function of the density $\rho(r, t)$:
\begin{equation}
      \Gamma_r(r, \textit{l}, t)  = - \frac{4 \pi G}{r^2} \int_0^r {\mathrm{d}}r r^2 \rho (r, t) +  \frac{\textit{l}}{r^3}
\textrm{.}
 \label{Eq:vlasnewsphergamma}
	 \end{equation}
This requires the knowledge of $\rho(r, t) $ in function of $f(r, u, \textit{l}, t)$. The number density $\rho(r, t) $ is the integral on the velocities of the position-velocity distribution (equation (\ref{eq:integra})). In the present case, this integral has to be done on a function of the variables $u$, radial velocity, and $\textit{l}$, squared angular momentum. Let us split the velocity into its component in the direction of $\bf{r}$, namely $u$, and its component perpendicular to  $\bf{r}$, say $\bf{w}$. This vector in a two dimensional space (the plane orthogonal to $\bf{r}$) is equal to the angular momentum, rotated by an angle of $\pi/2$ and divided by $r$. Moreover the distribution function depends on the length of $\bf{w}$ only. Therefore one can write the element of integration for $\bf{w}$ as $$ {\mathrm{d}}{\bf{w}} = \frac{2 \pi}{r^2} L {\mathrm{d}} L =  \frac{\pi}{r^2} {\mathrm{d}} \textit{l}\textrm{,}$$ This assumes isotropy in the plane perpendicular to $\bf{r}$, which explains the $2\pi$ factor coming from the angular integration on all directions in this plane.  Therefore the density $\rho(r, t) $ is given in function of $f(r, u, \textit{l}, t)$ as:
\begin{equation}
      \rho(r, t)  = \frac{\pi}{r^2}  \int_{-\infty}^{+\infty}{\mathrm{d}}u \int_0^{+\infty}{\mathrm{d}} \textit{l} f(r, u, \textit{l}, t)
\textrm{.}
 \label{Eq:rhofunctf}
	 \end{equation}
This completes the writing of the Vlasov-Newton set of equations in the spherically symmetric case, it includes Equations (\ref{Eq:vlasnewspher}), (\ref{Eq:vlasnewsphergamma}) and (\ref{Eq:rhofunctf}). Notice that this set of equation has an invariant, which can be seen as the probability distribution of the square momentum. This is
\begin{equation}
    g(\textit{l}) = \int_{-\infty}^{+\infty}{\mathrm{d}}u \int_0^{+\infty}{\mathrm{d}} r  f(r, u, \textit{l}, t)
\textrm{.}
 \label{Eq:rhofunctg}
	 \end{equation}
By integrating the equation of motion (\ref{Eq:vlasnewspher}) with respect to $u$ and $r$ one finds that $g(\textit{l})$ does not depend on time. One can consider models where the dependence of $g(.)$ with respect to $\textit{l}$ is simple, for instance
\begin{equation}
g(\textit{l}) = k_g \delta( \textit{l} - \textit{l}_0)
 \textrm{,}
 \label{Eq:glo}
	 \end{equation}
 where $\textit{l}_0$  is a given squared angular momentum and $k_g$ a normalising quantity. With this choice the steady solution of Equation (\ref{Eq:vlasnewspher}) are functions of the energy $$E' = \frac{u^2}{2} + \Phi(r) + \frac{\textit{l}_0}{2 r^2} \textrm{.}$$ This eliminate to have recourse to the (usual) assumption of independence of the steady distribution with respect to the squared angular momentum.
%  an independence which cannot be stated in terms of the initial conditions.
% je supprimerai la fin de cette phrase, si tu es d'accord ??
Such probability distribution of angular momentum with $l_0 \neq 0$ cannot yield a cusp at the origin in a steady state, because, as shown by the expression of $E'$, at short distance the term proportional to $\textit{l}$ depends on $r$ like $1/r^2$, although in a cusp-like solution, the potential $\Phi(r)$ increases at most like $r^{-1/2}$ as $r$ tends to zero.
In summary a cusp-core steady solution cannot have a distribution of angular momentum peaked at $\textit{l}_0 \neq 0$, but may exist for $\textit{l}_0 = 0$, a very peculiar case.
%Therefore such solutions with a density cusp at $r =0$ must have initially a distribution of angular momentum peaked at $\textit{l} =0$. Otherwise the conservation of angular momentum forbids the convergence of masses toward $r=0$.

\section{ An example of finite time singularity:  collapse of the dust gas}
\label{sec:collapsedust}

Having posed at hand the problem of the long time dynamics of solutions of Vlasov-Newton equation, we discuss now finite time dynamics.
%Of course, once a problem of dynamics like the one at hand is posed (like the one of the long time behavior of solutions of Vlasov-Newton equation),
One would like to know if a well defined solution exists at finite time. At the time of Michel H\'enon thesis, the numerical methods cannot solve more than about a hundred body problem. So, one had recourse to the analytical approach. In the present case the analytical results are limited, and established in the frame of  various assumptions which may or may not be realistic when compared to the result of accurate modern numerics. The only instance where there is a way of solving explicitly this set of equations is by assuming the spherical symmetry and that  initially there is no velocity at all, that amounts to assume a zero angular momentum too, as written above.  This defines the problem of the collapse of a dust gas (namely a gas without internal pressure, a \textit{free fall} process). Because of the absence of angular momentum, equation (\ref{eq:newt}) is transformed into
 \begin{equation}
   \ddot{r} = - \frac{G M(a)}{r^2} \mathrm{,}
   \label{eq:newt1}
   \end{equation}
where one has introduced the quantity $M(a)$, independent on time, and $a = r(0)$. This is a constant because the mass is carried by the motion. This is correct if the motion is in one direction only, toward the center. Otherwise one cannot relate in this way the mass in the interval $[0, r]$ to the one in $[0, a]$ at time zero.

This equation can be solved by introducing a parameter $\theta$ such that
 \begin{equation}
 r = a\cos^2(\theta)  \mathrm{,}
 \label{eq:rMestel}
   \end{equation}

and
 \begin{equation}
 t = \sqrt{\frac{3}{8 \pi G \overline{\rho}(a)}} (\theta + \frac{1}{2}\sin(2\theta))  \mathrm{.}
 \label{eq:tMestel}
   \end{equation}

where $\overline{\rho}(a) = \frac{3 M(a)}{4 \pi a^3}$, $\rho(a)$ being the density distribution, a function of the radius, at $t = 0$.

There is a singularity in the distribution of mass when $\theta$ becomes equal to $\pi/2$. Then the derivative $\frac{\partial t}{\partial \theta} $ becomes zero and so the mapping of $\theta$ into $t$ becomes singular. This happens first (as a time increases) when the prefactor of $(\theta + \frac{1}{2}\sin(2\theta))$ in equation (\ref{eq:tMestel}) is the smallest, which is for $a = 0$ if $\overline{\rho}(a)$ is maximum at $a = 0$.
%because then $r$ becomes zero for non zero values of $a$. This happens for the first time where $\overline{\rho}(a)$ is the largest, because it is in the denominator of the right-hand side of Equation (\ref{eq:tMestel}) giving $t$ as a function of $\theta$ and $a$.
This critical time for the occurrence of a singularity in the solution is $$t_c =  \frac{\pi}{2} \sqrt{\frac{3}{8 \pi G \overline{\rho}(0)}} \mathrm{.}$$ Near this critical time one can carry, as was done by Larson and by Penston \cite{Penston}, a local expansion of $\rho(r, t)$ for $t$ a slightly before $t_c$ and near $r =0$. As was not pointed out by those authors, the singular solution
%type of singularity
depends on the expansion of $M(a)$ near $ a = 0$. Suppose that $\rho(a)$ has a Taylor expansion $\rho(a) = \rho_0 + \rho_k a^k + ...$ with $\rho_k < 0$ and $k$ positive. It happens that the exponents of the singularity depend on the power $k$. If $k = 2$ as assumed by Larson and by Penston, one finds a certain set of exponents, but another set is found for $k =4$. This remark is of some relevance because we found \cite{Mannev} that such a free-fall solution described by equations (\ref{eq:rMestel})-(\ref{eq:tMestel}) appears at the end of the collapse of a sphere of \textit{compressible fluid}. This occurs for a model describing a kind of "soft compressibility", with a pressure increasing relatively slowly at large densities. In that case  gravitation forces becomes dominant in the late stage of this collapse, just before the singularity. Therefore, the late stage of this collapse is described by the same equations as the ones of the collapse of a dust gas. But it happens that the exponent are the ones corresponding to $ k = 4$, not $k = 2$ which could look more "natural". We showed that $k=4$ is chosen  because the asymptotic behavior of the core solution  can be matched with a free-fall solution decaying slowly far from the center, although the case $ k =2$ somehow requires a supersonic flow far from the center which  is not realized in general in a compressible fluid.

Compared to the fluid case, where the velocity and density field are single defined functions of $r$ and $t$, a dust gas does not have this constraint: flux of particles cross each other without creating any shock wave. Therefore it makes sense for this case to continue the solution after the time $t_c$ of occurrence of the singularity. In particular, it makes sense to  consider the question of the ultimate state reached at time infinity. Somehow this makes a prototypical problem of evolution of a gas of point masses interacting gravitationally only.

The simplest assumption one can make for the behavior of the dust gas is that particles cross $r= 0$ by keeping their kinetic energy and linear momentum. In the case of spherical symmetry, this condition is simple to state: the limit values of $f(r, u, t)$ at $r = 0$ are the same, for $u$ positive (outward motion) and negative (inward motion). Note that it would be more complicated to write the same condition in the non axis-symmetric case because one would have to make a distinction between different directions of space. The condition for conservation of energy and linear momentum writes

 \begin{equation}
 \lim_{r \to 0} f(r, u_+ , t) =   \lim_{r \to 0} f(r, u_-  = -u_+, t) \mathrm{,}
    \label{eq:fplus}
   \end{equation}
   where $u_+$ is positive and $u_-$ negative. This condition of continuity makes sense if the two functions $f(0, u_{\pm} , t)$ are finite. Otherwise, one has to replace it by the condition that the ratio of the two functions  $f(r, u_+ , t) $ and $f(r, - u_+ , t)$ tends to one as $r$ tends to zero. In principle this condition allows to continue the dynamics beyond time $t_c$.  This is made quite complex because, one cannot in general keep track of the ordering with respect of the center of the shells initially at different radii. Nevertheless it is possible to solve the problem of evolution at least for short times after the first singularity. We plan to return to this subject in the future.

However there is  simple case where one can continue the dynamics after the first singularity and even forever, the case where $\rho(a)$ is constant, not depending on $a$, initial radial distance. In this case all the dust sphere collapse at the same time on the center, this time being the one formerly denoted as $t_c$ which becomes independent on $a$ when $\rho$ is constant. What happens next is rather straightforward: all particles reaching the center at $t = t_c$ bounce back by following a reflected trajectory: their order in the radial direction is unchanged, as the fastest ones are those coming from the edge of the dust sphere. Therefore the mass $M(a)$ in equation (\ref{eq:newt1}) remains the same as it was before the singularity, the only difference with the dynamics before this singularity is that the radial velocities are now positive, directed outward. At time $2 t_c$ all particles reach a motionless state at the same radial position as they had at time $ t =0$, the initial time for the beginning of the collapse. Therefore, the motion of the whole dust sphere is periodic of period $2 t_c$, with a complete stop at each period and a collapse to the center a time $t_c$ after this complete stop. Indeed one expects this will be changed as soon as the initial density distribution $\rho(a)$ is not a constant anymore.

An understanding, even limited, of the problem of the solution of the dynamics of the dust sphere in this case would be very valuable, because it would yield a sure understanding of what happens in the dynamics of masses interacting gravitationally.
\section{Summary and conclusion}
\label{sec:summconc}
This contribution to the volume in tribute to Michel H\'enon has been written in a perhaps slightly unusual way, considering a specific problem, the occurrence of a finite time singularity of density, or density cusp. As we have seen this occurrence requires a number of special conditions to be satisfied, the most stringent from the point of view of physics is likely the absence, or at least the smallness of angular momentum in the initial conditions. As usual in Astrophysics, this question of the initial conditions is very hard to answer. The case of the dust gas (with no kinetic energy at the beginning) shows at least that such a singularity is possible.  Globular clusters, on the time scale of the period of the motion in the average field of the cluster should be seen as steady objects. However, and as had been seen by H\'enon, stars inside the cluster tend  to evaporate and one may wonder why such clusters are still there. It has been suggested \cite{Kaz} that the observed clusters are the result of an imbalance between evaporation and aggregation of surrounding stars in the Galaxy. They would be therefore more like dynamical than static equilibria, this being seen on the time scale of the (rare) binary encounters in the cluster which can put stars of the cluster on an escape trajectory or stars from the Galaxy in a bound trajectory. Therefore, if this is correct, the question of what happens on the "short" orbital time in the evolution of the cluster is unrelated to any physical stage in the history of real clusters.

As had been suggested for a rather long time it could be that "dark matter" is necessary to account for the rotation speed of Galaxies. This speed has long been observed to be noticeably larger than expected from the balance between the centrifugal force of visible matter and the centripetal inertia force. If this dark matter is made of particles interacting with their kins or with the others by gravitation only, one might wonder how this halo of dark matter survived over time the evaporation following close encounters either with the same particles of dark matter or ordinary matter in the Galaxy. This puts bounds on the properties of this dark matter, on their density and mass in particular.

Michel H\'enon was a provider of many new ideas in Astrophysics and in other fields as well. This communication tried to show how bright were his beginnings in Science and how inspiring they continue to be.

{}
 \ifx\mainismaster\UnDef%
 \end{document}